\documentclass[prb,twocolumn,showpacs,superscriptaddress,floatfix]{revtex4}

\usepackage{graphicx}
\usepackage[centertags]{amsmath}

\newcommand{\cm}{\ensuremath{\,\mbox{cm}^{-1}}}
\newcommand{\K}{\ensuremath{\,\mbox{K}}}
\newcommand{\celsius}{\ensuremath{\,{}^\circ}\!C}

\begin{document}

\title{Strong spin-phonon coupling in infrared and Raman spectra of SrMnO$_{3}$}

\author{S.~Kamba}\email{kamba@fzu.cz} \affiliation{Institute of Physics ASCR, Na Slovance~2, 182
21 Prague~8, Czech Republic}
\author{V.~Goian} \affiliation{Institute of Physics ASCR, Na
Slovance~2, 182 21 Prague~8, Czech Republic}
\author{V.~Skoromets}
\affiliation{Institute of Physics ASCR, Na Slovance~2, 182 21 Prague~8, Czech Republic}
\author{J.~Hejtm\'{a}nek}
\affiliation{Institute of Physics ASCR, Na Slovance~2, 182 21 Prague~8, Czech Republic}
\author{V.~Bovtun}
\affiliation{Institute of Physics ASCR, Na Slovance~2, 182 21 Prague~8, Czech Republic}
\author{M.~Kempa}
\affiliation{Institute of Physics ASCR, Na Slovance~2, 182 21 Prague~8, Czech Republic}
\author{F.~Borodavka}
\affiliation{Institute of Physics ASCR, Na Slovance~2, 182 21 Prague~8, Czech Republic}
\author{P.~Van\v{e}k}
\affiliation{Institute of Physics ASCR, Na Slovance~2, 182 21 Prague~8, Czech Republic}
\author{A.A.~Belik}
\affiliation{International Center for Materials Nanoarchitectonics (WPI-MANA), National
Institute for Materials Science (NIMS), 1-1 Namiki, Tsukuba, Ibaraki 305-0044, Japan}
\author{ J.H.~Lee} \affiliation{Department of Physics and Astronomy, Rutgers University, Piscataway, New Jersey
08854-8019, USA}
\author{ O.~Pacherov\'{a}}
\affiliation{Institute of Physics ASCR, Na Slovance~2, 182 21 Prague~8, Czech Republic}
\author{ K.M.~Rabe} \affiliation{Department of Physics and Astronomy, Rutgers University, Piscataway, New Jersey
08854-8019, USA}

\date{\today}

\pacs{75.80.+q; 78.30.-j; 63.20.-e}

\begin{abstract}

Infrared reflectivity spectra of cubic SrMnO$_{3}$ ceramics reveal 18\% stiffening of the
lowest-frequency phonon below the antiferromagnetic phase transition occurring at T$_{N}$
= 233\,K. Such a large temperature change of the polar phonon frequency is extraordinary
and we attribute it to an exceptionally strong spin-phonon coupling in this material.
This is consistent with our prediction from first principles calculations. Moreover,
polar phonons become Raman active below T$_{N}$, although their activation is forbidden
by symmetry in $Pm\bar{3}m$ space group. This gives evidence that the cubic $Pm\bar{3}m$
symmetry is locally broken below T$_{N}$ due to a strong magnetoelectric coupling.
Multiphonon and multimagnon scattering is also observed in Raman spectra. Microwave and
THz permittivity is strongly influenced by hopping electronic conductivity, which is
caused by small non-stoichiometry of the sample. Thermoelectric measurements show
room-temperature concentration of free carriers $n_{e}=$3.6 10$^{20}$ cm$^{-3}$ and the
sample composition Sr$^{2+}$Mn$_{0.98}^{4+}$Mn$_{0.02}^{3+}$O$_{2.99}^{2-}$. The
conductivity exhibits very unusual temperature behavior: THz conductivity increases on
cooling, while the static conductivity markedly decreases on cooling. We attribute this
to different conductivity of the ceramic grains and grain boundaries.

\end{abstract}

\maketitle

\section{Introduction}

In the last decade, there has been an intensive search for new multiferroic materials, in
particular those exhibiting ferroelectric (FE) and ferromagnetic (FM) order. The aim is
to find multiferroics with a large magnetoelectric coupling, which could allow the
control of magnetization by external electric field. Such materials could be used in
future electronic devices including non-volatile memories and spin filters.
Unfortunately, most multiferroics exhibit antiferromagnetic (AFM) order with a weak
magnetoelectric coupling or low critical temperatures, and thus they are not suitable for
practical applications.

One mechanism that has been demonstrated to produce novel multiferroics is application of
an epitaxial strain (or a comparable perturbation) to an AFM paraelectric material with a
large spin-phonon coupling such that the lowest frequency polar phonon with hypothetical
ferromagnetic (FM) ordering has lower frequency than the corresponding phonon in the AFM
phase. Lowering of the polar phonon frequency by the strain can produce a polar
instability in the FM phase, with the resulting energy lowering, which stabilizes a FE-FM
multiferroic phase over the bulk AFM paraelectric phase.\cite{Fennie06} This has been
experimentally confirmed in EuTiO$_3$ with a magnetic ordering temperature of
5.3\,K.\cite{Lee10a}

Recently, Lee and Rabe showed using first-principles calculations that this mechanism
would produce a multiferroic phase in SrMnO$_{3}$.\cite{Lee-Rabe10} Since bulk
SrMnO$_{3}$ is AFM below T$_{N}$=230\,K,\cite{Chmaissem01} strained thin films should be
FM with critical temperatures above 100 or 150\,K.\cite{Lee-Rabe10} However, due to the
very high critical strain (3.6\%), experimental efforts to realize this phase have been
so far unsuccessful. Tuning of the polar phonons by a related but different perturbation,
negative chemical pressure through continuous substitution of Sr by Ba, has recently been
investigated. Sakai \textit{et al.}\cite{Sakai11,Sakai12} observed ferroelectricity below
410 K in Sr$_{1-x}$Ba$_{x}$MnO$_{3}$ single crystals with $0.45 \leq x \leq 0.50$. The
polar distortion is found to be dominated by displacement of magnetic Mn$^{4+}$ cations
(running counter to the conventional expectation that $d$ electrons inhibit off-centering
of the cations \cite{Hill00}) and therefore the magnetoelectric coupling is the largest
attained so far.\cite{Sakai11} However, the magnetic ordering for the ferroelectric phase
is G-type AFM, so that ferroelectricity is the result of destabilization of the polar
mode by strain within the original AFM phase, and the system is not a true (FM-FE)
multiferroic.

The prospects for stabilizing a true multiferroic in SrMnO$_3$ or related compounds hinge
on better knowledge and control of the spin-phonon coupling. Fortunately, the spin-phonon
coupling can be directly investigated experimentally. Specifically, this coupling should
result in a remarkable stiffening of the Slater-type (i.e. vibration of Mn cations
against oxygen octahedra) mode frequency near and below T$_{N}$.\cite{Lee-Rabe11} Indeed,
unusually large shifts in phonon frequency at T$_{N}$ have recently been observed in
Sr$_{1-x}$Ba$_x$MnO$_3$ for x = 0-0.3,\cite{Sakai12} explaining the observed reduction in
the FE distortion in the AFM phase. In this paper, we investigate the spin-phonon
coupling and related properties through detailed THz, far infrared (FIR) and Raman
spectroscopic studies of SrMnO$_{3}$ and combine them with thorough investigations of
transport and dielectric properties.


\section{Experimental details}

The tolerance factor of SrMnO$_3$ is very close to unity ($\tau$=1.05) and therefore it
can be synthesized in both cubic and hexagonal crystal structures depending on the
growing procedure.\cite{negas70} We initially prepared the hexagonal SrMnO$_{3}$ powder
from a stoichiometric mixture of SrCO$_{3}$ and Mn$_{2}$O$_{3}$ by annealing in air at
1150\,\celsius\, for 50 h and at 1000\,\celsius\, for 24\,h. Then we pelletized the
hexagonal SrMnO$_{3}$ with a small amount of PVA and annealed at 1000\,\celsius\, for 55
h. Afterwards, pellets of the single-phase cubic SrMnO$_{3}$ were prepared in two
steps:\cite{Belik11} First, we annealed the pellets of hexagonal 4H-SrMnO$_{3}$ at
1500\,\celsius\, for 8 hours in an Ar flow with the heating/cooling rate of
300\,\celsius/min. The weight loss after the first step gave the composition of
SrMnO$_{2.53(2)}$. During the second step, we oxidized the obtained SrMnO$_{2.53}$ sample
in air at 350\,\celsius\ for 15 h. The weight gain after the second step gave the
composition of SrMnO$_{2.99(2)}$. The density of the ceramics was measured to be 5.50(2)
g/cm$^{3}$; it is about 97.0 \% of the powder density, and about 95.8 \% of the
theoretical XRD density.

The phase purity of the cubic SrMnO$_3$ ceramics was checked by x-ray diffraction and no
contamination by the hexagonal phase or by other second phases was discovered. We used
the diffractometer Bruker D8 DISCOVER equipped with rotating Cu anode
[$\lambda$(Cu$K\alpha1$)=1.540598 \AA; $\lambda$(Cu$K\alpha2$)=1.544426 \AA] operating
with 12\,kW power. A parabolic G\"{o}bel mirror on the side of the incident beam and
Soller collimator and 200-LiF analyzator on the side of the diffracted beam were used.
The crystal structure was measured between 170 and 303\,K. The temperature was controlled
by a cooling stage Anton Paar DCS 350 and no change of the crystal structure with
temperature was observed. The lattice parameter $a_{0}$ determined from 220 and 310 Bragg
diffractions linearly changed with temperature from 3.7988(3) at 170\,K to 3.8062(4)
\AA\, at 303\,K. We have to stress here, that we observed no anomaly in $a_{0}$(T) around
T$_{N}$.

\begin{figure}
  \begin{center}
    \includegraphics[width=60mm]{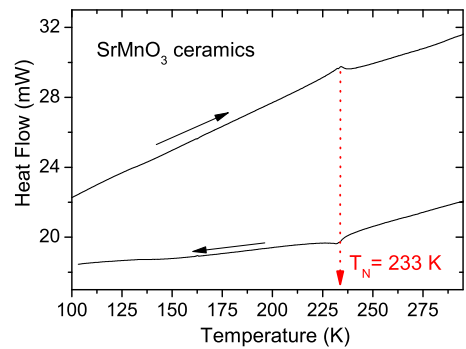}
  \end{center}
    \caption{(Color online) Heat flow measured on cooling and heating. It shows an anomaly
    near the antiferromagnetic phase transition at T$_{N}$=233\K\, without any thermal hysteresis.}
    \label{Fig1}
\end{figure}

The temperature of the magnetic phase transition was checked using calorimetry. The ceramic
pellet was placed in an aluminum pan and measured using a differential scanning
calorimeter Perkin Elmer Pyris Diamond DSC. The sample was cooled from 300\,K down to
100\,K with a cooling rate of 10\K/min. The same rate was used for heating. An anomaly at
233\,K corresponding to the AFM phase transition was observed in both cooling and heating
curves without any temperature hysteresis (see Fig.~\ref{Fig1}).

The spectroscopic experiments were performed using a Fourier-transform infrared (IR)
spectrometer Bruker IFS\,113v equipped with a helium-cooled bolometer (operating
temperature 1.6\,K) as a detector and a custom-made time-domain THz
spectrometer.\cite{kuzel10} In both experiments, Optistat CF cryostats (Oxford
Instruments) with polyethylene (far IR) or Mylar (THz) windows were used for measurements
in the 10 -- 300\,K temperature range. In the THz spectrometer, a femtosecond Ti:sapphire
laser oscillator (Coherent, Mira) produces a train of femtosecond pulses which generates
linearly polarized broadband THz pulses in a photoconducting switch TeraSED
(Giga-Optics). A gated detection scheme based on electro-optic sampling with a
1\,mm-thick [110] ZnTe crystal as a sensor allows us to measure the time profile of the
electric field of the transmitted THz pulse (see Ref. \onlinecite{kuzel10} for further
details).

Ceramic pellets were carefully polished using diamond paste in order to obtain a flat
surface with an optical quality. For IR studies, the sample was glued on a circular
metallic aperture with a diameter of 5\,mm. The sample IR reflectivity was normalized by
the reflectivity of a gold mirror sputtered on polished glass, which was glued on the
similar aperture. An optical cryostat allowed us to move the sample holder at all
temperatures, so we have measured the sample and reference spectra at each temperature.
For the time-domain THz transmission experiments we prepared a polished plane-parallel
plate of ceramics with thickness of 50\,$\mu$m.

IR reflectivity and THz complex permittivity spectra were carefully fitted assuming the
dielectric function in the form of sum of classical three-parameter damped oscillators
\cite{gervais83}
\begin{equation}
\label{eps3p}
 \varepsilon^*(\omega)
 = \varepsilon_{\infty} + \sum_{j=1}^{n}
\frac{\Delta\varepsilon_{j}\omega_{TOj}^{2}} {\omega_{TOj}^{2} -
\omega^2+\textrm{i}\omega\gamma_{TOj}} \, ,
\end{equation}
where $\omega_{TOj}$, $\gamma$$_{TOj}$ and $\Delta\varepsilon_{j}$ denote the
eigenfrequency, damping and dielectric strength of the $j$-th transversal optic (TO)
polar phonon. $\varepsilon$$^{*}$($\omega$) is related to the reflectivity R($\omega$) of
the bulk sample by
\begin{equation}\label{refl}
R(\omega)=\left|\frac{\sqrt{\varepsilon^{*}(\omega)}-1}{\sqrt{\varepsilon^{*}(\omega)}+1}\right|^2
.
\end{equation}
The high-frequency permittivity $\varepsilon_{\infty}$ = 7.6 given by electronic
absorption processes was obtained from the room-temperature frequency-independent
reflectivity tails above the phonon frequencies. The temperature dependence of this
quantity was neglected, consistent with its behavior in other related perovskite
dielectrics.\cite{Roessle13}

For Raman studies, a Renishaw RM 1000 Micro-Raman spectrometer equipped with a CCD
detector and a Linkam THMS 600 temperature cell was used. The experiments were performed
in a backscattering geometry in the 100 -- 2000\,cm$^{-1}$ range. He-Ne laser with the
wavelength 633\,nm was used.

The magnetic susceptibility data were collected with a Quantum Design (MPMS)
superconducting quantum interference device when the applied magnetic field was 100\,Oe. The
thermoelectric power and thermal conductivity measurements were carried out between 3.5
and 320\,K using a close-cycle refrigerator. The four-point steady-state method was
applied in order to eliminate thermal resistances between sink and heater; the electrical
contacts were realized using a silver paint. The experimental setup was checked using
reference samples. For more details we refer to Ref. \cite{Hejtmanek02}.

Dielectric measurements in the frequency range from 1 MHz to 1.8 GHz were performed using
a computer-controlled dielectric spectrometer equipped with an Agilent 4291B Impedance
analyzer, Novocontrol BDS 2100 coaxial sample cell, and Sigma System M18 temperature
chamber (operating range of 100 -- 500\,K). The dielectric parameters were calculated
taking into account the electromagnetic field distribution in the sample.

\section{Calculational details}

First-principles calculations were performed using density-functional theory within the
generalized gradient approximation GGA + $U$ method with the Perdew-Becke-Erzenhof
parametrization as implemented in the Vienna ab initio simulation package (VASP-4.6).
Details are described in Ref. \onlinecite{Lee-Rabe11}. The phonon frequencies depend on
the choice of the lattice constant and the value of onsite Coulomb interaction $U_{\rm
eff}$. With $U_{\rm eff}$ = 2.3\,eV, the zero-temperature phonon frequencies computed for
antiferromagnetic spin ordering at $a_0$ = 3.825 \AA\, intermediate between the computed
lattice constant and the theoretical lattice constant, are in good agreement with the
low-temperature experimental values. We note a previous first-principles calculation of
phonons at the experimental lattice constant $a_0$ = 3.806 \AA\, yielded good agreement
for the TO2 frequency\cite{Sondena07}; however, this is not inconsistent as the latter
calculations did not include spin polarization. The temperature dependent spin
correlation ($\langle$S$_i$$\cdot$S$_j$$\rangle$) was calculated using mean-field theory
with the exchange constant computed at $a_0$ = 3.825 \AA\, and from this the temperature
dependent phonon frequencies are obtained using the spin-phonon parameterization of Ref.
\onlinecite{Fennie06}.

\begin{figure}
  \begin{center}
    \includegraphics[width=58mm]{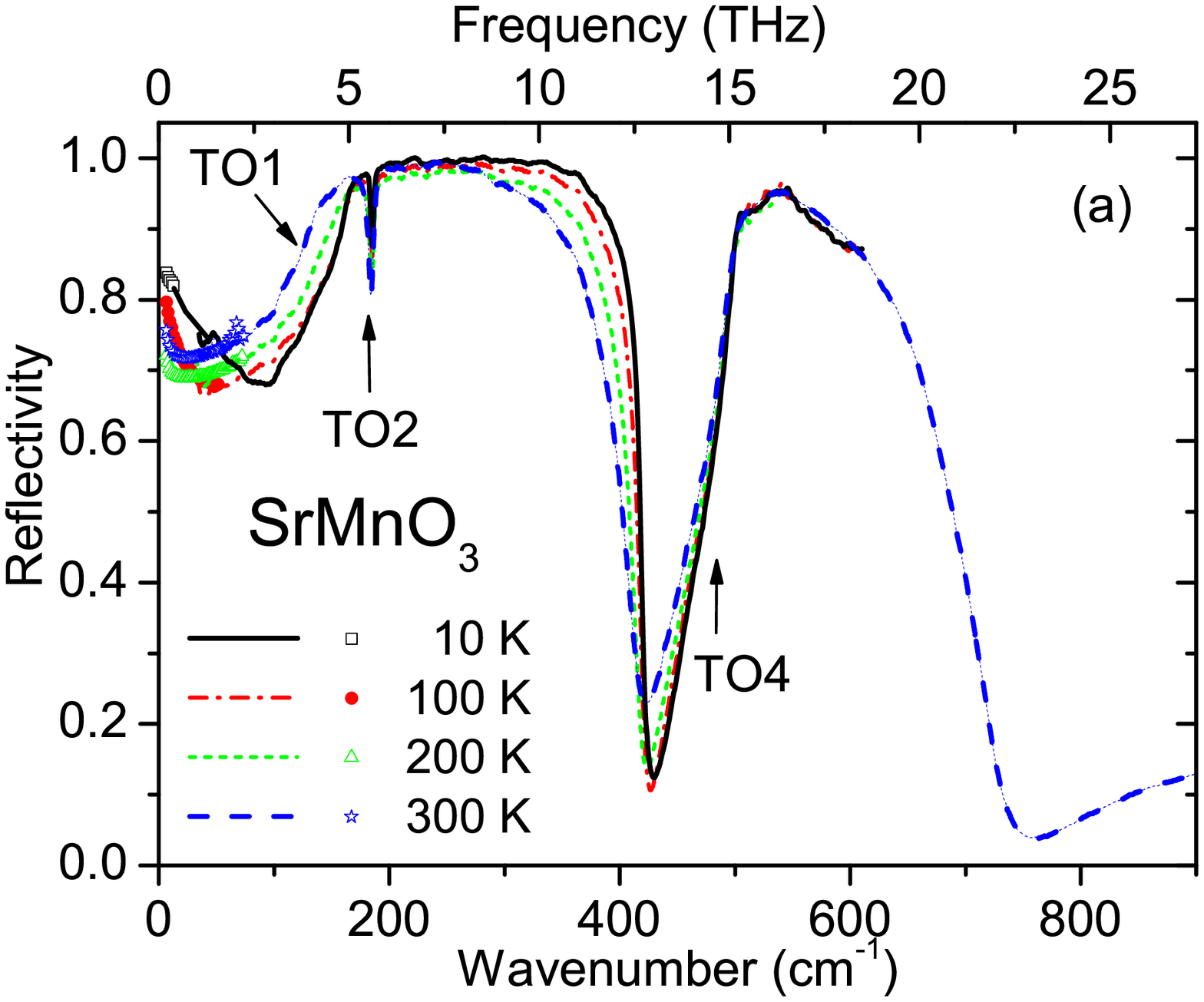}
    \includegraphics[width=65mm]{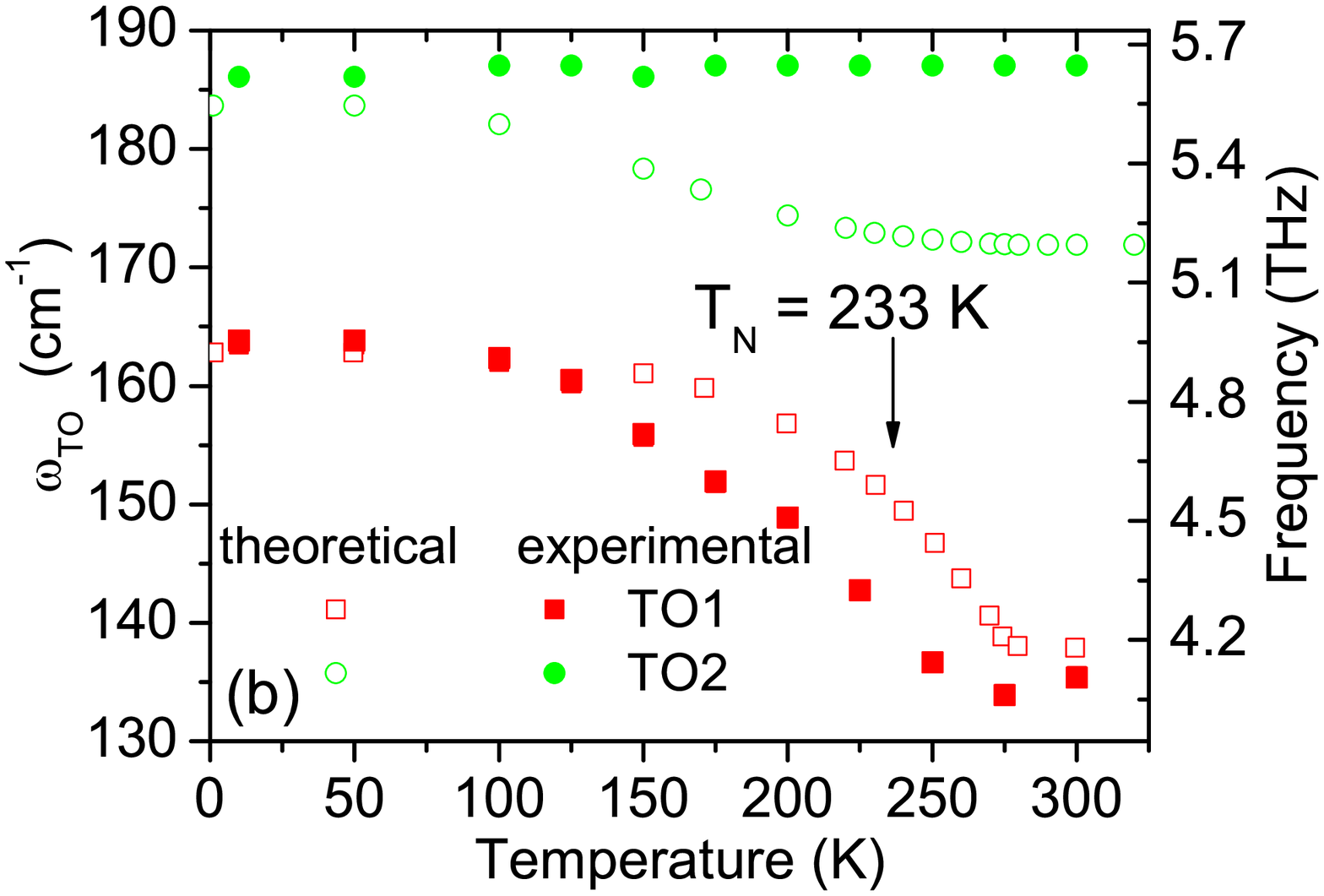}
    \includegraphics[width=65mm]{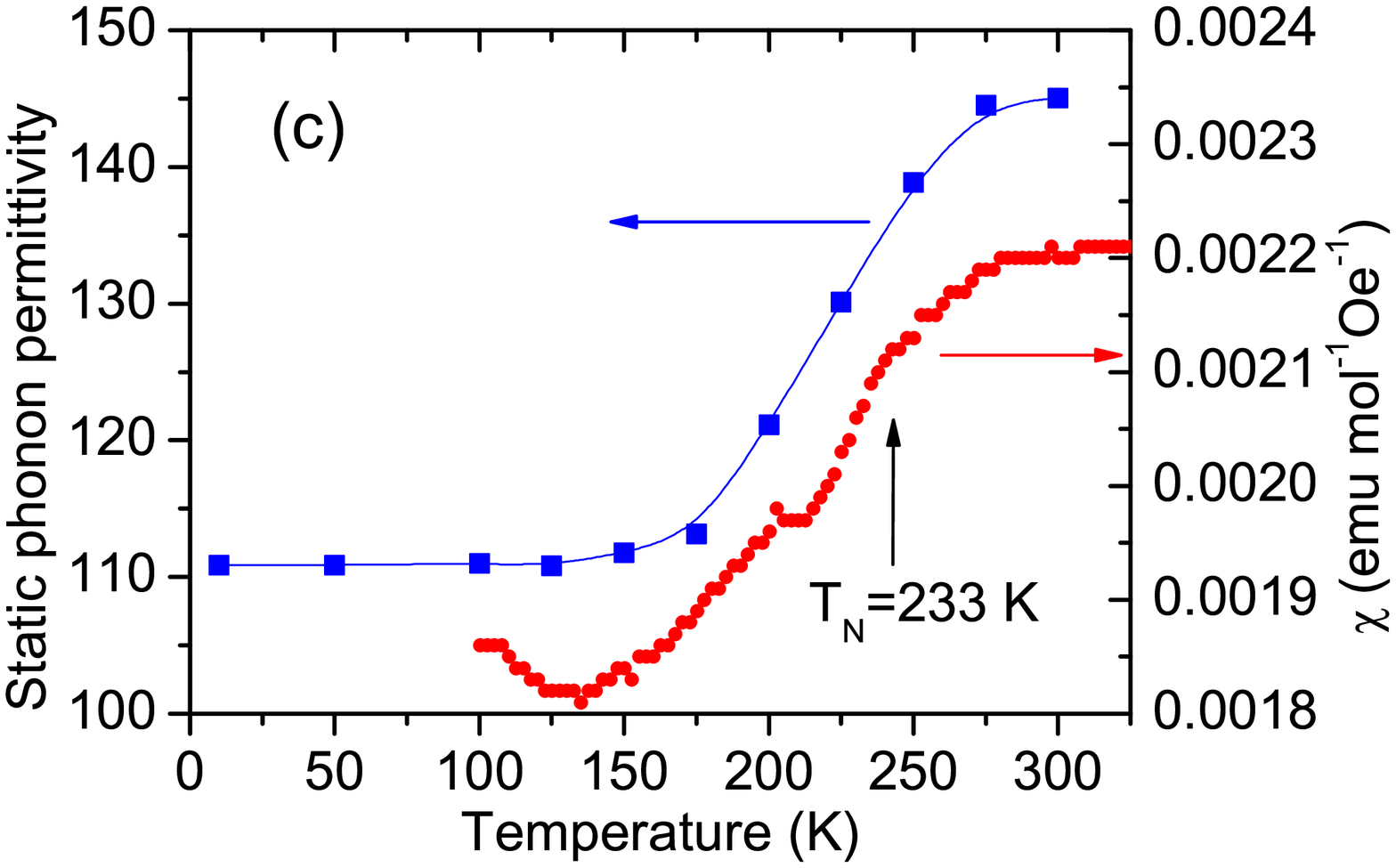}
  \end{center}
    \caption{(Color online) (a) Infrared reflectivity spectra at various temperatures. Frequencies
    of all three polar phonons are marked by arrows (TO3 is silent).
    (b) Temperature dependence of TO1 and TO2 phonon frequencies obtained from the fits of IR reflectivity
     (solid symbols) and first principles calculations
    (open symbols). (c) Temperature dependence of the static phonon permittivity obtained from the
    sum of all phonon and
  electronic ($\varepsilon_{\infty}$=7.6) contributions (left scale). Temperature dependence of
  the static magnetic susceptibility (right scale). }
    \label{Fig2}
\end{figure}

\section{Results}

\subsection{IR spectrocopy studies}

\begin{figure}
  \begin{center}
   \includegraphics[width=55mm]{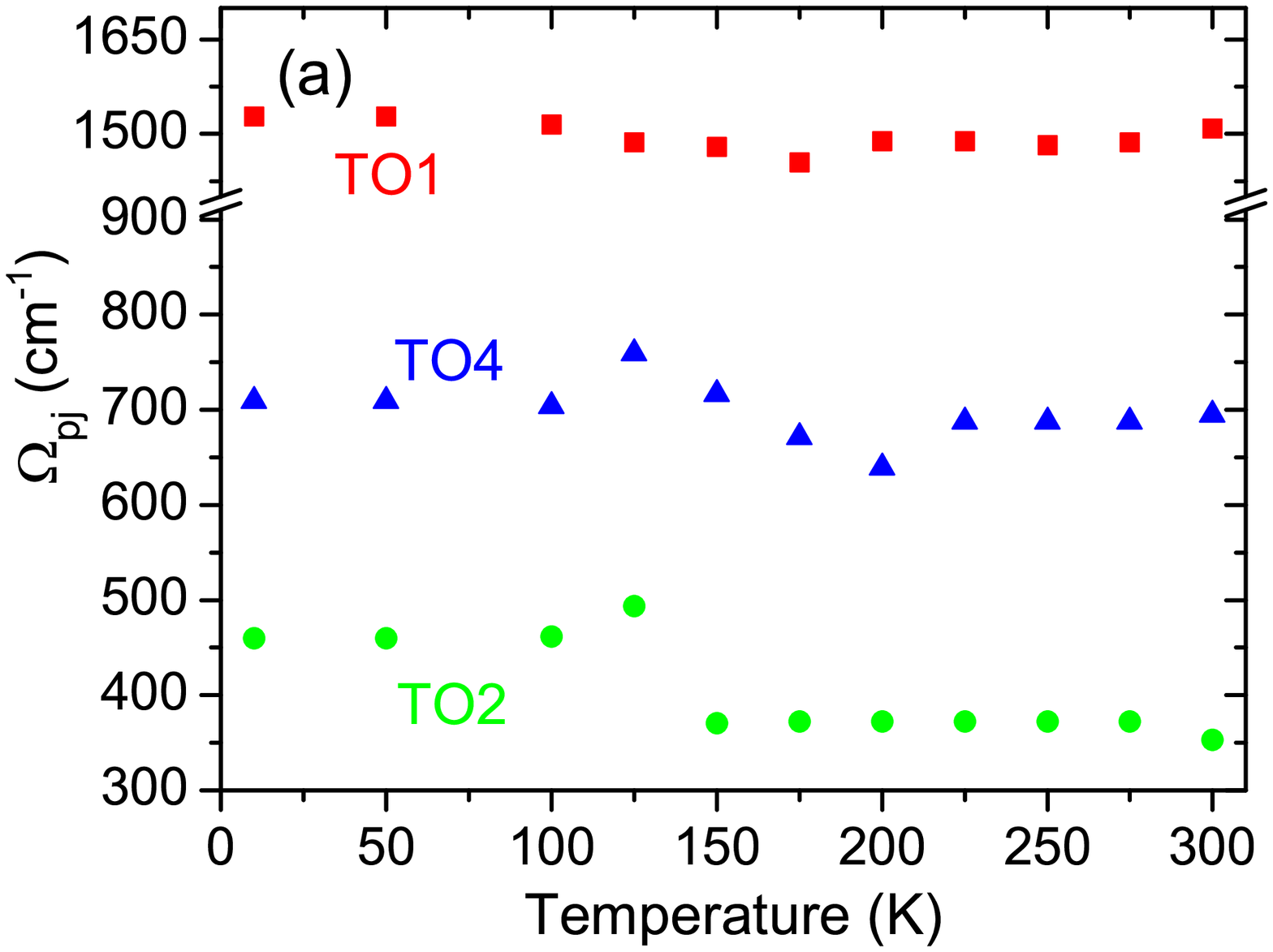}
    \includegraphics[width=55mm]{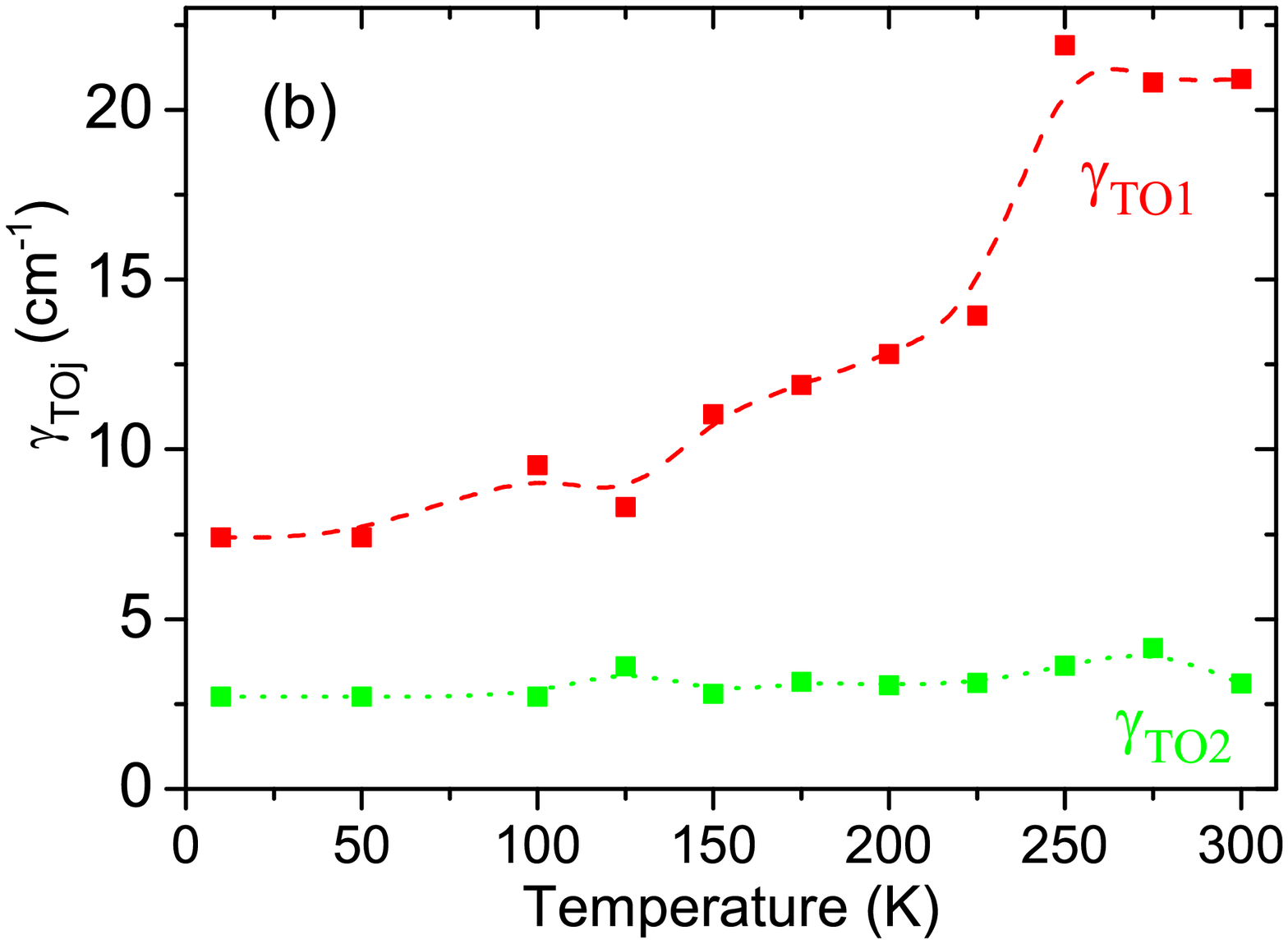}
  \end{center}
    \caption{(Color online) (a) Temperature dependence of experimental
    plasma frequencies $\Omega_{pj}=\sqrt{\Delta\varepsilon_{j}}\omega_{TOj}$ of the all polar phonons. (b)
    Temperature dependence of TO1 and TO2 phonon
    damping obtained from the fits of IR spectra. Lines are guides for eyes.}
    \label{Fig3}
\end{figure}

FIR reflectivity spectra of cubic SrMnO$_{3}$ taken at various temperatures down to 10\,K
are shown in Fig.~\ref{Fig2}a. Three polar optical modes (symmetry $F_{1u}$) typical of the
cubic $Pm\bar{3}m$ perovskite structure are seen. The lowest-frequency TO1 phonon
undergoes a remarkable temperature-induced shift of its frequency $\omega_{TO1}$.
It increases by 18\% on cooling (Fig.~\ref{Fig2}b) and therefore its dielectric strength
$\Delta\varepsilon_{1}$ decreases by 42\% (from 124 to 87) with decreasing temperature.
The dielectric strength of the other two phonons is small
($\Delta\varepsilon_{2}+\Delta\varepsilon_{4}=16$) and temperature independent; therefore
the static phonon permittivity $\varepsilon(0)=\sum
\Delta\varepsilon_{j}+\varepsilon_{\infty}$ exhibits decrease by 32\% on cooling (see
Fig.~\ref{Fig2}c). The change in $\Delta\varepsilon_{j}$ follows from the requirement that the
plasma frequency $\Omega_{pj}=\sqrt{\Delta\varepsilon_{j}}\omega_{TOj}$ of the $j$-th
uncoupled phonon should be temperature independent. Indeed, examination of the spectra shows that all the plasma frequencies are temperature independent within experimental
error (see Fig.~\ref{Fig3}a). Sakai $et\,al.$\cite{Sakai12} observed different behavior
in SrMnO$_{3}$ single crystal, namely decrease of $\Omega_{p1}$ and simultaneous increase
of $\Omega_{p2}$. It means that the plasma frequency transfers from the TO1 to the TO2
mode, which is a signature of their mutual coupling. This mode coupling is caused by the
fact that $\omega_{TO1}$=180\,\cm\, and $\omega_{TO2}$=190\,\cm\, are very close in
single crystals at low temperatures, while in the investigated ceramics we found TO1 mode
to be less stiffened ($\omega_{TO1}$=164\,\cm) and therefore it is less coupled to TO2.
Nevertheless, we observed a large enhancement of TO1 phonon damping near and above
T$_{N}$ (see Fig.~\ref{Fig3}b). The discrepancy of $\omega_{TO1}$ in the ceramics and
crystal can be attributed to different oxygen content in both samples. Note, we annealed the
ceramics at 350\,\celsius\, in air and at ambient pressure, while the crystal was
annealed at 480\,\celsius\, in oxygen atmosphere at pressure of 6.5\,GPa.\cite{Sakai11}

The temperature dependence of the static permittivity $\varepsilon'$(0) obtained from IR
spectra is shown in Fig.~\ref{Fig2}c. $\varepsilon'$(0) consists of the sum of dielectric
contributions of all three polar phonons and electrons. The decrease of $\varepsilon'$(0)
by 32\% on cooling below T$_{N}$ follows from Lydanne-Sachs-Teller relation
$\frac{\varepsilon'(0)}{\varepsilon_{\infty}}=\prod
\frac{\omega_{LOj}^{2}}{\omega_{TOj}^{2}}$ due to the $\omega_{TO1}$ stiffening on
cooling (longitudinal phonon frequency $\omega_{LOj}$ is usually temperature
independent). Note that the temperature change of $\varepsilon'$(0) is one order of
magnitude larger than that found in EuTiO$_{3}$,\cite{Katsufuji01} which is assumed to be
a material with very strong spin-phonon coupling. Note as well that the static magnetic
susceptibility $\chi$ exhibits temperature dependence similar to that of $\varepsilon'$(0) (see
Fig.~\ref{Fig2}c). We have measured $\chi$(T) up to 500\,K and from its fit by a
Curie-Wess law we have obtained the Curie-Weiss temperature $\Theta$ $\cong$ -2300\,K.
This gives evidence about strong magnetic correlations above T$_{N}$. Similar $\chi$(T)
behavior was reported by Belik \textit{et al.}\cite{Belik11}

Note that $\chi$(T) starts to decrease already near 270 K (Fig.~\ref{Fig2}c), 40\,K above T$_{N}$.
TO1 phonon frequency exhibits a minimum as well near 270\,K
(Fig.~\ref{Fig2}c). This can be caused by short-range magnetic correlations above
T$_{N}$, hints of which are also seen in the temperature dependence of the thermal conductivity
$\kappa$ (see discussion in Section C.)

The experimentally observed TO1 phonon frequency and its temperature dependence
correspond very well to the theoretical prediction obtained from first-principles
calculations (see Fig.~\ref{Fig2}b). Our calculations show that the TO1 phonon
eigenvector corresponds to a Slater-type vibration, i.e. to the Mn-vibrations against
oxygen octahedra. In Fig.~\ref{Fig3}a we see that this mode has the highest plasma
frequency $\Omega_{P1}$, which gives experimental evidence that the TO1 phonon is the
Slater mode.\cite{Hlinka06} The theoretical TO2 mode frequency corresponds well to the
experimental value at 10 K, but unlike the experimental mode, it softens slightly with
increasing temperature and at room temperature it is 10\% lower than the experimental
value. However, this behavior is sensitive to the choice of lattice constant and the
temperature dependence of the TO2 mode is reduced in calculations for which the chosen
lattice constant is increased. Note that $\omega_{TO2}$ is experimentally determined very
precisely ($\pm$1\cm), because its damping is very low (see Figs.~\ref{Fig2}a and
~\ref{Fig3}b). A temperature independent $\omega_{TO2}$(T) like that shown in
Fig.~\ref{Fig2}b was reported by Sakai \textit{et al.}.\cite{Sakai12}

Sacchetti \textit{et al.}\cite{Sacchetti05} investigated FIR reflectivity spectra of
cubic and hexagonal SrMnO$_{3}$ at 100 and 300\,K and obtained the same result as found
here, but they did not present detailed temperature dependence of the phonon frequencies.
Nevertheless, they have shown that 4H and 6L hexagonal phases of SrMnO$_{3}$ have 12 and
14 IR active modes, respectively.\cite{Sacchetti05} We found only 3 polar phonons in the
whole experimental temperature range. This confirms that our SrMnO$_{3}$ ceramics is not
contaminated by any detectable secondary hexagonal phase.

\subsection{Raman spectroscopy studies}

\begin{figure}
  \begin{center}
    \includegraphics[width=60mm]{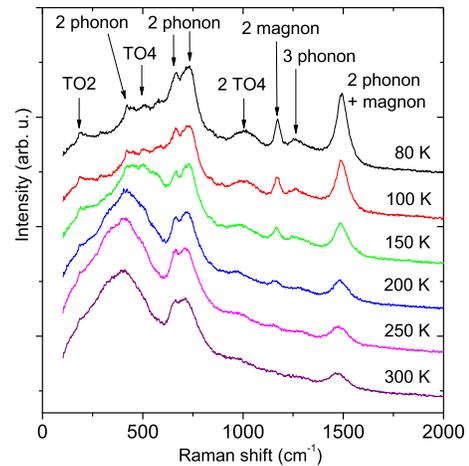}
  \end{center}
    \caption{(Color online) Micro-Raman spectra of SrMnO$_{3}$ ceramics taken at various temperatures.
    The peaks seen at 80\,K are assigned.}
    \label{Fig4}
\end{figure}

\begin{figure}
  \begin{center}
    \includegraphics[width=70mm]{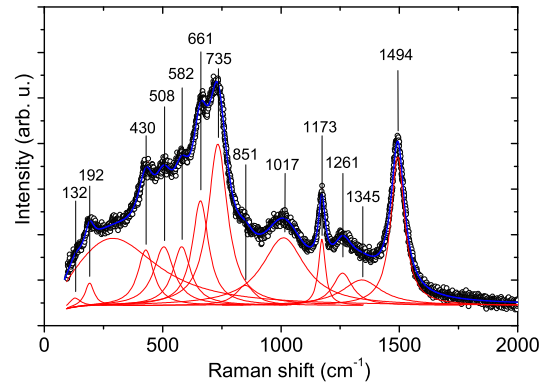}
  \end{center}
    \caption{(Color online) Raman scattering spectrum taken at 80\,K and its decomposition into
    independent damped harmonic oscillators. Frequencies of the Raman modes are marked.}
    \label{Fig5}
\end{figure}

Micro-Raman spectra were measured down to 80\,K. SrMnO$_{3}$ crystallizes in the cubic
$Pm\bar{3}m$ structure, where the all phonons should be Raman inactive and only 3
$F_{1u}$ symmetry modes are IR active.\cite{Hlinka06} In spite of this fact, three broad
bands are seen in the room-temperature Raman spectrum (see Fig.~\ref{Fig4}). These bands
come, as they do in SrTiO$_{3}$,\cite{Perry67} from Raman scattering of higher order, i.
e. from multiphonon summation processes. The bands split on cooling, because the phonon
damping decreases towards low temperatures. Several new modes also appear outside of the
region of the original three bands. The experimental Raman spectrum measured at 80\,K and
its fit with assigned mode frequencies are shown in Fig.~\ref{Fig5}. The frequencies of
the peaks at 192 and 508\,\cm\, correspond to TO2 and TO4 frequencies (in our assignment,
the TO3 mode is silent in IR and Raman spectra of perovskites with $Pm\bar{3}m$
structure). However, polar TO2 and TO4 modes are forbidden in Raman spectra of the cubic
structure. This means that the Raman scattering spectra give evidence of locally broken
cubic structure and center of symmetry below the magnetic phase transition temperature,
although our x-ray diffraction experiment performed down to 170\,K did not reveal any
change of the macroscopic symmetry. Similar magnetically induced changes of the phonon
selection rules were recently reported in various transition-metal monooxides and
chromium spinels.\cite{Kant12} In that work, structural investigations did not reveal any
changes. However, IR spectra resolved a remarkable phonon splitting below T$_{N}$. This
was explained by the exchange coupling which slightly distorts the local cubic symmetry
of the crystals and activates new phonons in the spectra.\cite{Kant12}

There are several other new Raman bands activated below T$_{N}$. The frequency of the
band seen at 1017\cm\, corresponds exactly to twice the TO4 mode frequency. A sharp band
appears below T$_{N}$ at 1173\cm. This is most probably a two magnon scattering coming
from magnons at the Brillouin zone boundary. Two other strong peaks are seen near 1300
and 1500\,\cm. These come from multiphonon and/or phonon-magnon scattering. In this
summation process contribute mainly phonons and magnons with maximal density of states,
i.e. with wavevector from Brillouin zone boundary. Similar multiphonon and multimagnon
Raman bands arising below T$_{N}$ were observed in BiFeO$_{3}$.\cite{Ramirez08,Ramirez09}

The newly activated Raman modes cannot come from a secondary hexagonal phase, because the
hexagonal phase does not have any phonons above 800\cm\, and our low-frequency modes also
do not correspond to phonons of hexagonal SrMnO$_{3}$. \cite{Sacchetti05,Sacchetti06}
Moreover, 4H hexagonal SrMnO$_{3}$ exhibits structural and magnetic phase transitions
already around 380\,K and 280\,K, respectively.\cite{Daoud07}

\subsection{Dielectric and conductivity studies}

\begin{figure}
  \begin{center}
    \includegraphics[width=60mm]{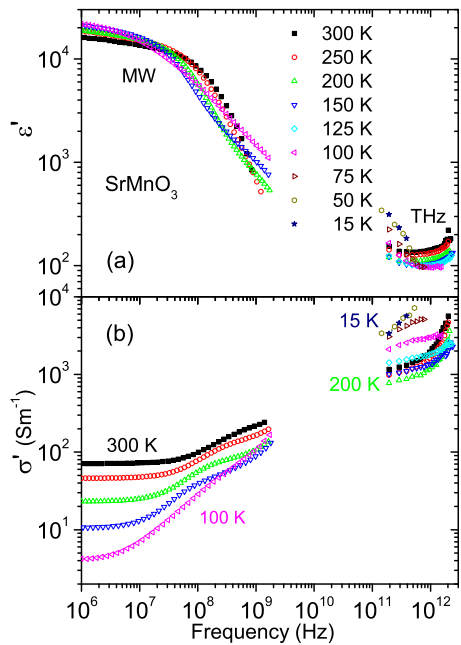}
  \end{center}
    \caption{(Color online) Frequency dependence of the microwave and THz (a) dielectric permittivity and (b)
    conductivity measured at various temperatures.
    Note that the MW data were obtained only above 100\,K.}
    \label{Fig6}
\end{figure}

The static permittivity $\varepsilon'$(0) shown in Fig.~\ref{Fig2}c is given by phonon
and electron contributions and it corresponds to THz experimental values measured between
100 and 300\,K. However, values of the megahertz permittivity are two orders of magnitude higher
(see Fig.~\ref{Fig6}a). This is caused by a non-homogeneous conductivity of the
sample in the ceramic grains and grain boundaries, which is responsible for the creation of
internal barrier layer capacitors on the grain boundaries. This mechanism is responsible
for a ``giant" effective permittivity in many dielectric or multiferroic materials.
\cite{Lunkenheimer02,Li09,Kamba07a}

Real part of the conductivity $\sigma'(\omega)$ exhibits strong frequency dependence
typical for hopping processes.\cite{Funke93} Its low-frequency part is almost frequency
independent and it corresponds to a static (\textit{DC}) conductivity of the percolated
grain boundaries. The high-frequency part of $\sigma'(\omega)$ increases with frequency.
However, it shows a plateau between 0.1 and 1 THz. Then $\sigma'(\omega)$ again increases
towards higher frequencies. The plateau THz conductivity value corresponds to the
conductivity in the ceramic grains, while the rising $\sigma'(\omega)$ above 1\,THz is
caused by a phonon absorption.

Unfortunately, we could not study grain size dependence of the conductivity, because we
could not prepare ceramics with various grain sizes. As mentioned already in section II,
we had to transform the initial hexagonal ceramics to the cubic phase by 8 hour annealing
at 1500\,\celsius\, followed by rapid quenching of the ceramics. The grains grow at such
high temperature that we could not control their size. We should mention that
Sakai's single crystal was also conducting, and therefore they published only 1\,GHz
permittivity at 4\,K. No radio-frequency or microwave data of SrMnO$_{3}$ taken at higher
temperatures were published by Sakai \textit{et al.} or by other authors.

\begin{figure}
  \begin{center}
    \includegraphics[width=70mm]{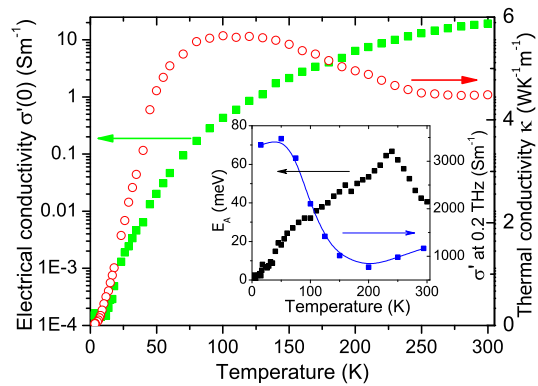}
  \end{center}
    \caption{(Color online) Temperature dependence of the static electrical conductivity $\sigma'(0)$ and
    thermal conductivity $\kappa$.
    Temperature dependence of the activation energy $E_{a}$ of the \textit{DC} conductivity  and conductivity
    $\sigma$' measured at 0.2 THz are shown in the inset.}
    \label{Fig7}
\end{figure}

\begin{figure}
  \begin{center}
    \includegraphics[width=60mm]{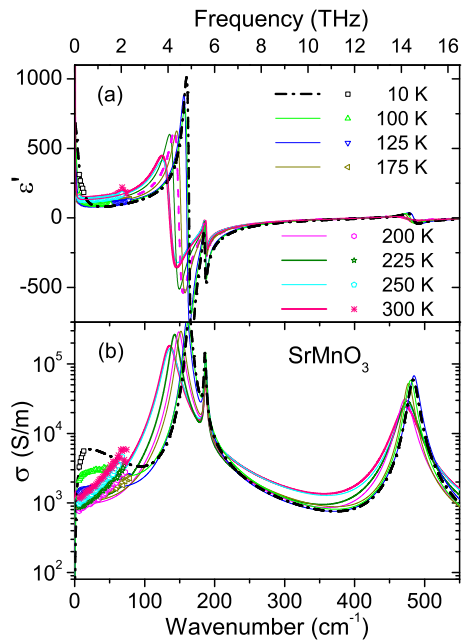}
  \end{center}
    \caption{(Color online) (a) Real part of the complex dielectric function and (b) optical conductivity
    in SrMnO$_{3}$
    obtained from the fits of the
    THz and IR spectra measured at various temperatures. At low frequencies and low temperatures arises
    a peak in the conductivity, which we fitted using highly damped oscillator.}
    \label{Fig8}
\end{figure}

The frequency dependent conductivity $\sigma'(\omega)$ exhibits a very interesting
temperature dependence (see Fig.~\ref{Fig6}b and inset of Fig.~\ref{Fig7}). The static
and low-frequency $\sigma'(\omega)$ exhibits classical semiconductor behavior, i.e. it
decreases on cooling. However, the THz $\sigma'(\omega)$ exhibits complex temperature
behavior. Its value decreases on cooling from 300\,K to 200\,K, but on further cooling it
increases and at 15\,K reaches almost one order of magnitude higher values than that at
200\,K. The former decrease of the THz conductivity on cooling is caused by a reduced
phonon damping at low temperatures and by a corresponding lower phonon absorption in the
THz region. However, on further cooling below 200\,K the mean free path of charge
carriers in the ceramic grains dramatically increases and therefore also the THz
$\sigma'(\omega)$ increases. It can be caused by so called magnon drag
effect,\cite{Zanmarchi68} when magnons promote the high-frequency conductivity. The THz
conductivity and permittivity spectra were possible to fit together with the IR
reflectivity spectra using a sum of damped oscillators from Eq.~\ref{eps3p}. The
high-frequency hopping conductivity is seen at low temperatures as a peak near 30\cm\, in
the optical conductivity $\sigma'(\omega)$ - see Fig.~\ref{Fig8}. The higher-frequency
peaks correspond to the polar phonons.

In order to determine accurate chemical composition, specifically the concentration of
oxygen vacancies and consequently of Mn$^{3+}$ species, we have measured the thermoelectric
power. Using the standard configuration entropy approach,\cite{Maignan02} we have
determined the concentration of Mn$^{3+}$ in SrMn$^{4+}$O$_{3}$ using the room
temperature thermoelectric power ($S\sim -350$\,$\mu$VK$^{-1}$) as 2\% of all Mn$^{3+}$
species. This results in the chemical formula of our ceramic sample
Sr$^{2+}$Mn$_{0.98}^{4+}$Mn$_{0.02}^{3+}$O$_{2.99}^{2-}$. The same composition was obtained
independently by
measurement of the sample weight changes due to oxidation during annealing.\cite{Belik11}

In addition, we have measured the temperature dependence of the thermal conductivity $\kappa$ down
to 4\,K. These data are shown together with the local activation energy $E_{a}$ of
$\sigma'$(0) in Fig.~\ref{Fig7}. Most typically, the thermal conductivity of a
nonmetallic material is dominated by the phononic part, which we assumed to be true for the case of
SrMnO$_{3}$. The temperature dependence of $\kappa$ is then governed by the specific
heat, sound velocity and phonon mean free path. These are decisively influenced by
various scattering mechanisms. Considering the Debye temperature $\Theta_{D}\sim 400$\,K
of SrMnO$_{3}$ we anticipate the ``classic" hyperbolic increase of $\kappa$ below room
temperature, reflecting the dissipative Umklapp phonon-phonon scattering. At low
temperatures this is followed by the decrease critically limited by the grain boundary
scattering. As we are dealing with a magnetic material, we should anticipate the phonon
scattering on magnons, which should reduce $\kappa$. This is in fact observed in
SrMnO$_{3}$; $\kappa$(T) is temperature independent down to $\sim$ $T_{N}$ and only a
very moderate increase with decreasing temperature to 100\,K is observed. Below this
temperature $\kappa$ decreases due to the decreasing specific heat. Let us note that
similar behavior was recently reported by Suzuki $\it{et\, al.}$ \cite{Suzuki12} who have
shown that the observed $\kappa$(T) is dominated by acoustic phonons. We conclude that
based on the temperature variation of $\kappa$ the phonons are critically scattered not
only by magnons below T$_{N}$, but also by paramagnons above T$_{N}$. Contrary to the
complex temperature dependence of $\kappa$, the static electrical conductivity
$\sigma'$(0) (Fig.~\ref{Fig7}) exhibits monotonous decrease from 20 Sm$^{-1}$ (at 300\,K)
down to $10^{-4}$ Sm$^{-1}$ without apparent anomalies. Nevertheless, detailed evaluation
of the temperature dependence of $\sigma'$(0) reveals a maximum of the local activation energy
$E_{a}$ near $T_{N}$, underlining the interrelation between the magnetic order and charge
carrier transport. The decrease of $E_{a}$(T) to zero suggests hopping as the most
likely mechanism of charge carrier transport. The fact that the $T_{N}$ is reflected by
the large, broad but distinct anomaly in electric transport gives evidence that the
charge, lattice and magnetic degrees of freedom are strongly coupled in SrMnO$_{3}$.

\section{Conclusion}

First principles calculations predicted a large stiffening of TO1 phonon frequency below
T$_{N}$, which we experimentally confirmed. In consequence, the static phonon
permittivity drops by 32 \%. Spin-phonon coupling also influences the change of the local
crystal symmetry, which allows activation of the polar phonons in Raman scattering
spectra. Moreover, multiphonon and multimagnon peaks in the Raman spectra appear, sharpen
and split on cooling. The MHz permittivity reaches giant values due to the conductivity
of the ceramic grains. It comes from a small nonstoichiometry of the ceramics with
chemical formula Sr$^{2+}$Mn$_{0.98}^{4+}$Mn$_{0.02}^{3+}$O$_{2.99}^{2-}$. The sample
non-stoichiometry originates from annealing at 1500\,\celsius\, during the ceramic
preparation and its subsequent fast quenching to room temperature. Unfortunately, this
procedure is necessary for stabilization of the cubic perovskite structure in
SrMnO$_{3}$, because the hexagonal structure of SrMnO$_{3}$ is otherwise more stable.
Sample annealing performed in air at 350\,\celsius\, significantly modifies the oxygen
content of the sample, which changes from SrMnO$_{2.53}$ to SrMnO$_{2.99}$, but
unfortunately it was not possible to obtain a fully stoichiometric sample. On one hand,
the static conductivity always decreases on cooling because it is influenced by the
semiconducting grain boundaries. On the other hand, the mean free path of charge carriers
in ceramic grains increases on cooling, and therefore also the THz conductivity increases
with lowering temperature.

Our data show exceptionally high spin-phonon coupling in
SrMnO$_{3}$, which could be used for induction of the ferroelectric phase in strained
thin films or doped bulk samples. Nevertheless, since good insulating behavior of the
samples is necessary for possible multiferroic applications, perfect stoichiometry of
SrMnO$_{3}$ is required. Annealing at a high hydrostatic pressure in oxygen atmosphere
could be a way to prepare highly resistive samples with strong
magnetoelectric coupling.

\begin{acknowledgments}

This work was supported by the Czech Science Foundation (Projects No. P204/12/1163 and
14-14122P) and M\v{S}MT (SIMUFER COST projects LH13048 and LD11035). The research was
also partially supported by WPI Initiative (MEXT, Japan), FIRST Program of JSPS, and the
Grants-in-Aid for Scientific Research (22246083) from JSPS (Japan).

\end{acknowledgments}


\end{document}